# THE KINEMATICS OF EMISSION-LINE GALAXIES IN CLUSTERS FROM ENACS

A. Biviano[1], P. Katgert[1], A. Mazure[2], M. Moles[3], R. den Hartog[1], P. Focardi[4]

[1] *Sterrewacht Leiden, The Netherlands*
[2] *Laboratoire d'Astronomie Spatiale, Marseille, France*
[3] *Instituto de Astrofísica de Andalucía, CSIC, Granada, Spain*
[4] *Dipartimento di Astronomia, Università di Bologna, Italy*

By using the ENACS data, we have studied the properties of emission-line galaxies (ELG) in clusters. The fraction of ELG in clusters is 16 %; this fraction is slightly dependent on the cluster velocity dispersion, being larger for "colder" systems.

ELG tend to occur more frequently than other cluster galaxies in substructures. However, only a minor cluster fraction has an ELG population with a mean velocity significantly different from that of the overall cluster population.

The velocity distribution of cluster ELG is broader than that of the other cluster galaxies, which translates in a larger observed velocity dispersion. The velocity distribution for ELG in the inner cluster regions suggests that ELG are on mostly radial orbits.

## 1. INTRODUCTION

Galaxies of different morphological types live in different environments (e.g. Hubble & Humason 1931): Dressler (1980a) clearly established the dependence between the fraction of early- and late-type cluster galaxies, and their local density. Early and late-type galaxies not only differ in their spatial distribution, but also in their kinematics, as first shown by Moss & Dickens (1977), and later confirmed by Sodré et al. (1989) and Biviano et al. (1992), for samples of 15 and 37 galaxy clusters, respectively. The segregation of morphological types with respect to their local density kinematics, indicates that either the formation process or the subsequent evolution of a galaxy or both are affected by the galaxy environment.

As a cluster probably forms through the collapse of a density perturbation with a radially decreasing density, a time sequence of infalling shells of galaxies is expected: the spirals would then be on infalling orbits while the ellipticals and S0's would constitute the virialized cluster population (Tully & Shaya 1984). The observations of a different mean velocity for early- and late-type galaxies in 3 out of 4 clusters, by Zabludoff & Franx (1993), and the numerical simulations by van Haarlem & van de Weygaert (1993), indicate that this infall could be anisotropic.

In this paper, we re-examine the issue of a different kinematical behaviour for early- and late-type galaxies, by using the extensive data-base on cluster galaxies provided by the ENACS (ESO Nearby Abell Cluster Survey). As morphologies are not available for all the galaxies in our survey, we identified two galaxy populations according to the the presence/absence of emission lines in their spectra. These two populations are referred as ELG (Emission-Line Galaxies) and, respectively, ALG (Absorption-Line Galaxies) in the following.

## 2. THE DATA-BASE

The ENACS data-base contains 5634 galaxies with reliable redshifts in the directions of 107 clusters from the catalogue of Abell, Corwin & Olowin (1989), with richness $R \geq 1$ and mean redhsifts $z \leq 0.1$. Magnitudes are also available for almost all these galaxies. Along these 107 line-of-sights, 220 systems with at least 4 galaxies are identified in the velocity space (see Katgert et al. 1995). Removing interlopers is possible only when the data-set is large enough; we have identified and rejected interlopers only in systems with at least 50 galaxies, by using the method described in den Hartog & Katgert (1995; see also Mazure et al. 1995).

In the wavelength and redshift ranges covered by our observations, the principal emission lines that can be detected are [OII] (3727 Å), H$\beta$ (4860 Å) and the [OIII] doublet (4959, 5007 Å). Emission lines have been identified independently by two of us (AB and MM) in two different ways, i.e. through a direct examination of the 2D Optopus-CCD frames, and by looking at the uncleaned wavelength-calibrated 1D spectra; 1231 ELG have been found. The good agreement between the absorption-line based redshifts and the emission-line based redshifts for the same galaxies, suggests that most of the identified emission-lines are real (see Katgert et al. 1995).

From a subsample of 548 galaxies in 10 ENACS clusters with available morphologies from the catalogue of Dressler (1980b, kindly provided by the author in electronic form), we find that 87 % of our ELG are spirals, but only 30 % of the Dressler's spirals are classified as ELG. Therefore, the properties of ELG should be regarded as typical of a subset of the spiral class, possibily biased towards larger equivalent widths of the emission lines.

In Fig. 1a we show the fraction of ELG – e.g. the number of ELG divided by the total number of galaxies – in systems with different numbers of galaxy members, N, and in the field – e.g. galaxies not assigned to any system –, vs. N. It can be seen that the ELG fraction is a decreasing function of N, from a value of $0.45 \pm 0.03$ for field galaxies (in good agreement with Zucca et al. 1995, these proceedings), to an asymptotic value of $0.16 \pm 0.01$ for systems with $N \geq 20$. If N traces the richness of a given system, this result is consistent with the well known dependence of the spiral fraction on the cluster richness (Oemler 1974; Bahcall 1977).

While systems with $N \geq 20$ appear to have a roughly constant ELG fraction, we show in Fig. 1b that among these systems, those with a larger $\sigma_v$ have a lower ELG fraction. This suggests that it is the *mass* of a system, rather than its richness, that is directly related to the ELG fraction, since $\sigma_v$ is a better mass tracer than N.

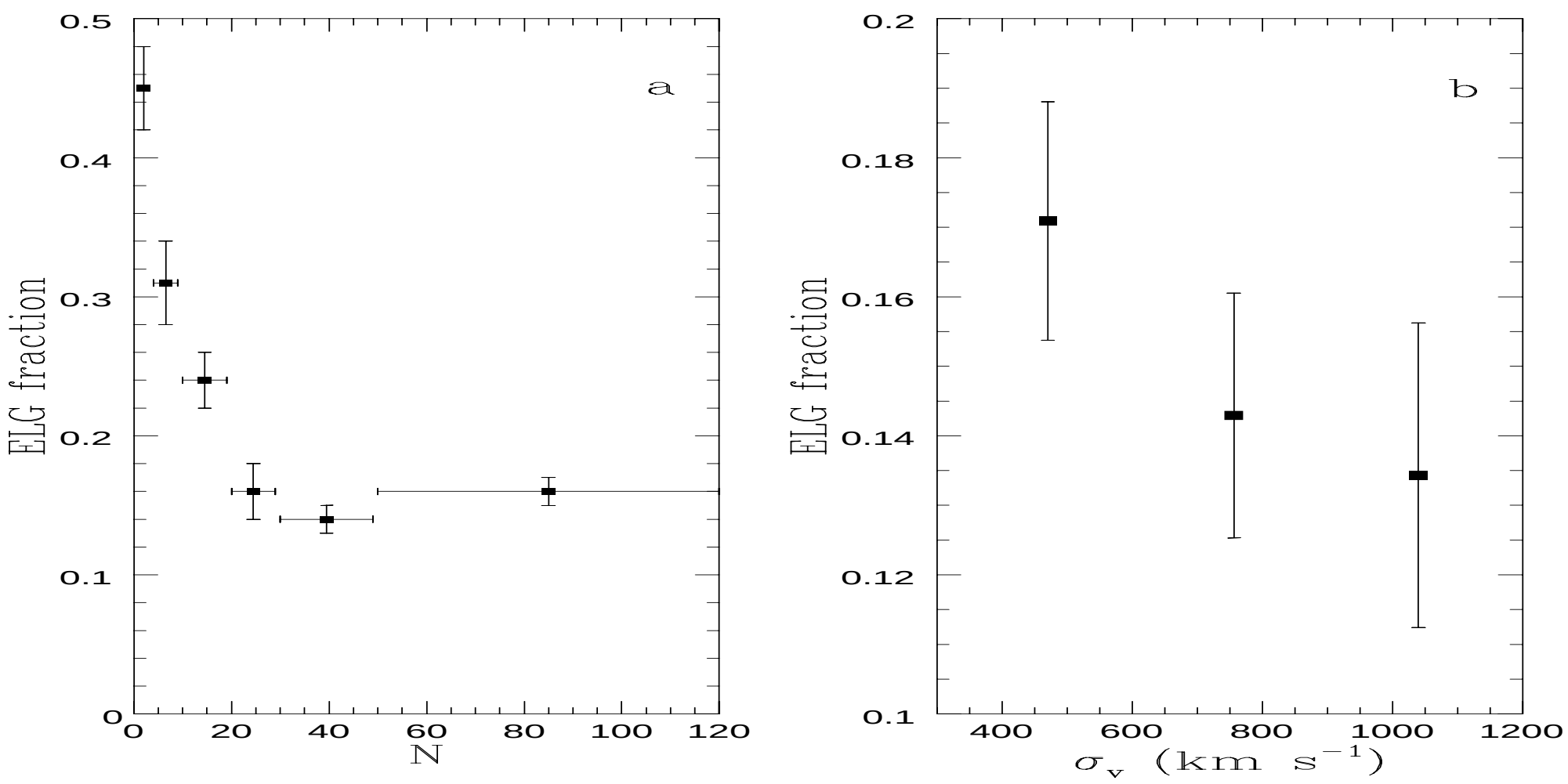

FIGURE 1. a) The fraction of ELG in the field, and in systems with different numbers of galaxy members, N. Poissonian error bars are shown on the vertical axis, and the ranges in the spanned N, on the horizontal axis. b) The average ELG fraction in systems with respect to their velocity dispersion, $\sigma_v$. 1-$\sigma$ error bars are shown on the vertical axis.

## 3. SUBSTRUCTURES

The findings of Zabludoff & Franx (1993) suggests that cluster ELG may be preferentially located within substructures. We have run a test for the difference of the <v>'s of ELG and ALG, based on the Student's $t$ statistic, on the 20 clusters with at least 10 ALG ($N_{ALG} \geq 10$) and at least 10 ELG ($N_{ELG} \geq 10$). The differences between the two mean velocities vs. the mean velocities of these 20 systems are shown in Fig. 2a. Only in three clusters the difference is significant (at $\geq 95$ % confidence level, c.l. hereafter); these clusters are labelled in the figure.

A more performing test for the detection of substructures is that of Dressler & Shectman (1988), modified according to the prescriptions of Bird (1994). We have run this test on the 25 clusters with $N \geq 50$, since these clusters have been cleared from interlopers, and since this test is unreliable when only few galaxies are available. In 8 clusters we find evidence for substructure at the 99 % c.l.; when excluding ELG from each cluster sample, only 6 clusters still show evidence for substructure at the same c.l. Moreover, the ELG fraction is larger in the 8 clusters with evidence of substructure: $0.22 \pm 0.02$ as opposed to $0.12 \pm 0.01$ in the remaining 17.

We are thus led to conclude that ELG are more often found in substructures than ALG, although this is unlikely to be a general rule for all clusters.

## 4. KINEMATICS

We have plotted in Fig. 2b the differences between the $\sigma_v$'s computed for ELG

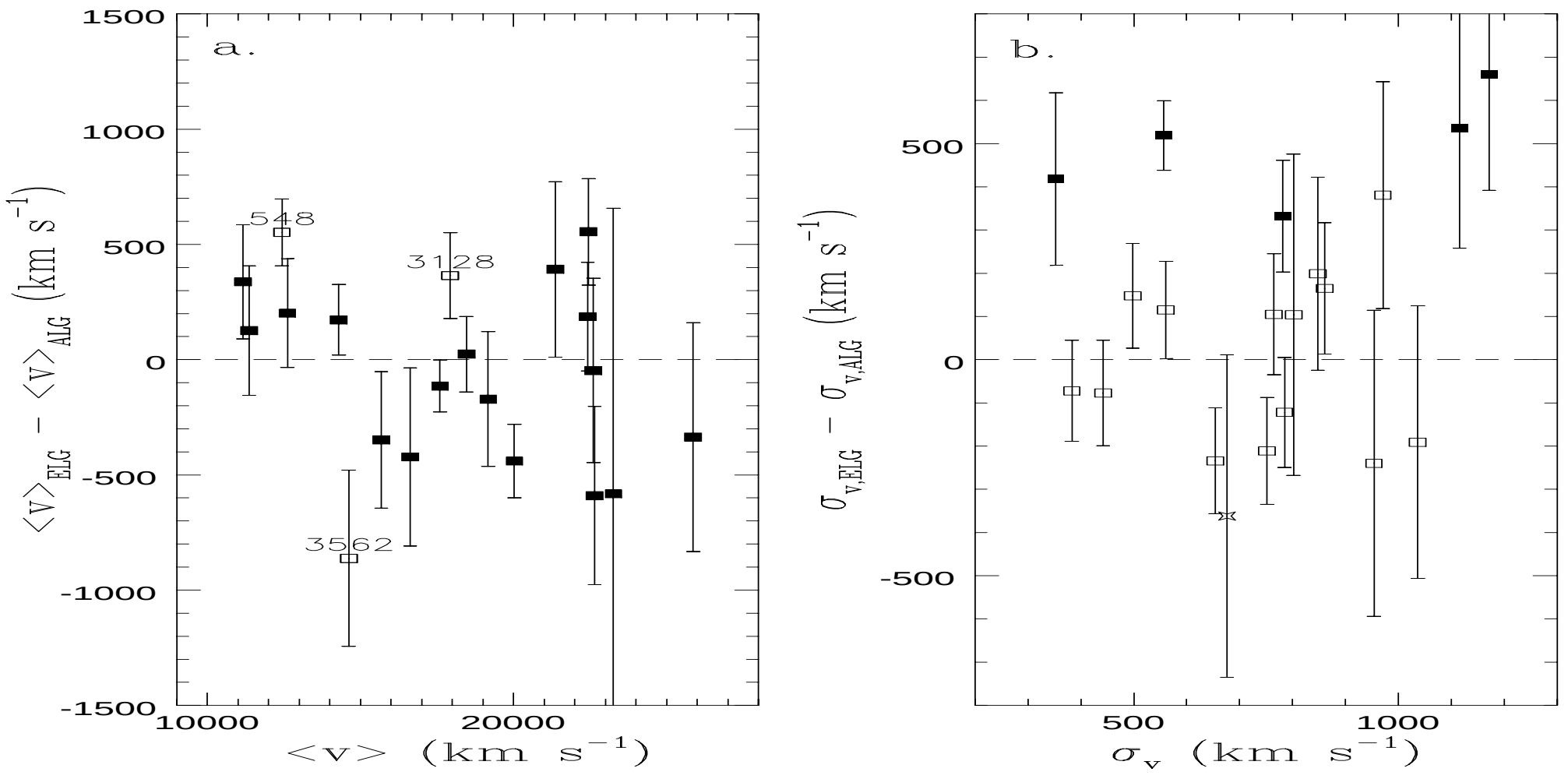

FIGURE 2. a) The difference between the average velocities of the ELG and the ALG populations in 20 clusters with $N_{ALG} \geq N_{ELG} \geq 10$, vs. the average velocities of the global population. The 1-$\sigma$ error bars on the differences are shown. The three clusters for which the velocity difference is $> 95$ % significant are indicated. b) The difference between the velocity dispersions of the ELG and the ALG populations in 20 clusters with $N_{ALG} \geq N_{ELG} \geq 10$, vs. the velocity dispersions of the global population. The 1-$\sigma$ error bars on the differences are shown. The filled squares and the star indicate those clusters with $\sigma_{v,ELG}$ significantly ($> 95$ %) larger and, respectively, lower, than $\sigma_{v,ALG}$.

and ALG for 20 clusters with $N_{ALG} \geq N_{ELG} \geq 10$, vs. the global $\sigma_v$'s of these clusters. The F-test for the difference of two variances indicates that there are 5 clusters in which the variance of the ELG velocities is significantly larger (c.l. $\geq$ 95 %) than the variance of the ALG velocities, and for only one cluster (Abell 3764) the opposite is true.

In Fig. 3, we show the two cumulative distributions of $\sigma_v$'s for 75 clusters with $N \geq 20$, as computed on all the galaxies, and respectively on ALG only. As it can be seen, removing ELG from the cluster samples lowers the value of $\sigma_v$; a Wilcoxon-test indicates that the probability for the two $\sigma_v$-distributions to be drawn from the same parent population is $< 0.001$.

Both the F- and the Wilcoxon-test indicate that the $\sigma_v$ for ELG tend to be larger than the $\sigma_v$ for ALG; however, since the fraction of ELG is only $\sim 0.16$, the *global* value of the cluster $\sigma_v$ is hardly affected.

To further examine the kinematical properties of ELG, we have put together all cluster galaxies in a single total sample. To this purpose, the galaxy velocities have been referred to the average velocity of the cluster which they belong to, and normalized to the cluster velocity dispersion, giving (v$-$ <v>)$/\sigma_v$. In order to have a robust estimate of <v> and $\sigma_v$, only clusters with $N \geq 20$ have been considered. There are 585 ELG and 3729 ALG in this total sample.

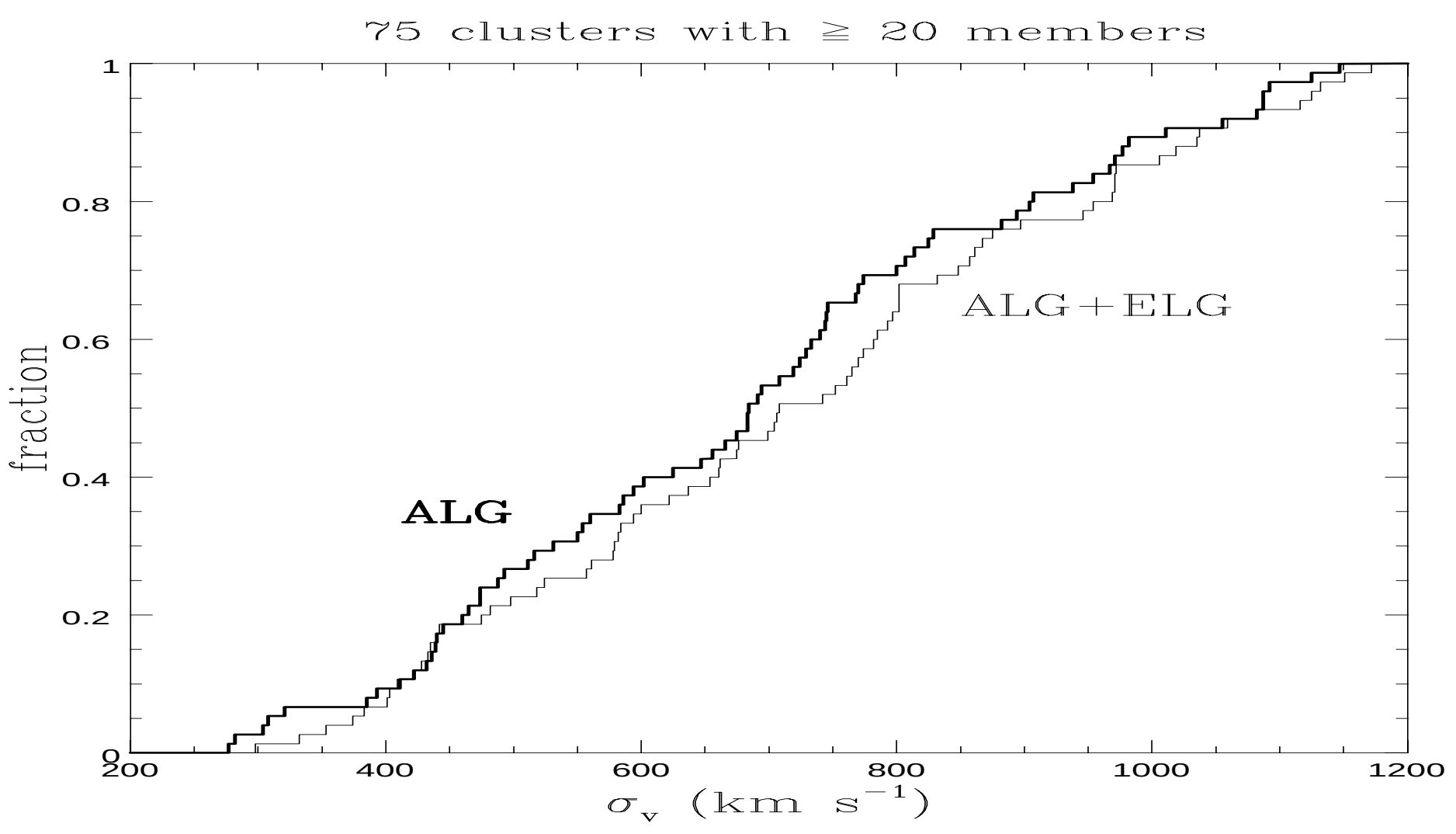

FIGURE 3. The cumulative $\sigma_v$ distributions for ALG only (thick line), and for all the galaxies (ALG+ELG, thin line) of 75 clusters with at least 20 members.

The ELG and ALG (v− <v>)/$\sigma_v$-distributions are plotted in Fig. 4a. The KS-test gives a probability of 0.035 that the two distributions are drawn from the same parent population. It can be seen that the distribution for ELG is broader than that for ALG; this is also indicated by the value of $1.23 \pm 0.05$ for the ratio between the average values of |v− <v> | /$\sigma_v$ for ELG and ALG.

Two effects may induce a broader velocity distribution: (a) a larger $\sigma_v$; (b) a difference in the <v>of ELG and ALG, since the velocities we use are referred to the <v>of the whole population, which is very nearly the same as the <v>of ALG, but not necessarily of ELG. Hypothesis (b) is unlikely to be correct, for the following reasons: (i) only a minority of clusters shows evidence for a different <v>for ELG and ALG; (ii) the normalized-velocity distributions of ALG and ELG in clusters without evidence for subclustering are still significantly different; (iii) the normalized-velocity distributions of ELG in clusters with and, respcetively, without evidence for subclustering are not significantly different. We conclude that most of the kinematical difference between ELG and ALG is due to the fact that the ELG velocity distribution has a larger dispersion, as found above for individual clusters (see § 3).

However, the dispersion does not fully characterize a distribution, when this is not gaussian, and the distribution of ELG normalized-velocities is different from a gaussian (99.4 % c.l., according to an Omnibus test). The deviation from gaussianity mostly comes from ELG in the inner cluster regions (see Fig. 4b), which have a platikurtic normalized-velocity distribution. Such a behaviour is expected if the ELG orbits are predominantly radial (see, e.g. Merritt & Saha 1993) A preliminary analysis shows that a very strong level of anisotropy is required if ELG have to be in equilibrium with the cluster potential, as traced by ALG.

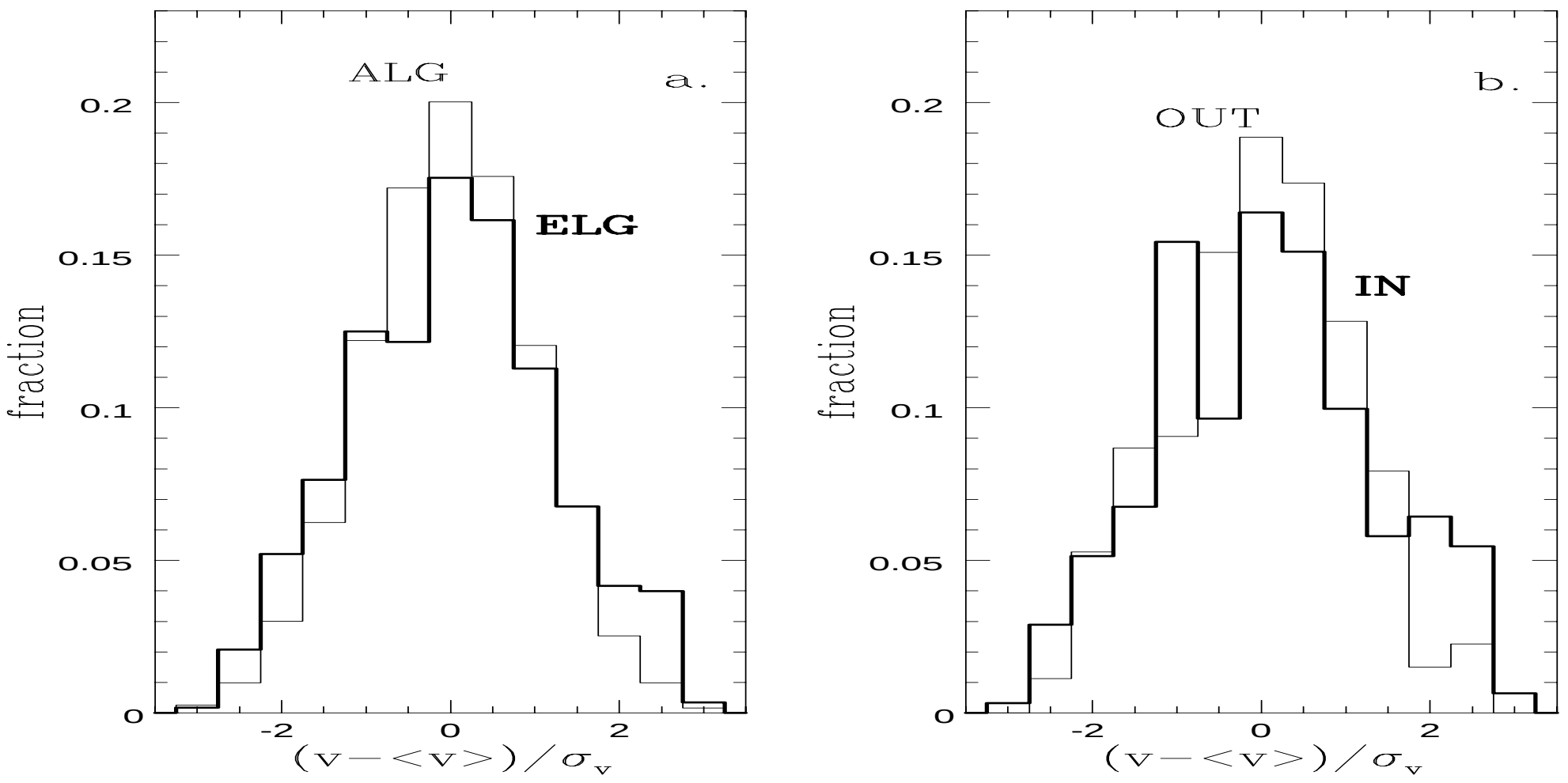

FIGURE 4. a) The normalized-velocity histograms for the total sample of 585 ELG (thick line) and 3729 ALG (thin line) in the 75 clusters with at least 20 members. b) The normalized-velocity histograms for ELG within (thick line) and, respectively, outside (thin line) 0.5 $h^{-1}$ $Mpc$ from their cluster centers.